\documentclass[sigconf, nonacm]{acmart}

\AtBeginDocument{
  \providecommand\BibTeX{{
    \normalfont B\kern-0.5em{\scshape i\kern-0.25em b}\kern-0.8em\TeX}}}

\begin{document}

\title{Local Perceptions and Practices of News Sharing and Fake News}

\author{Gionnieve Lim}
\email{gionnievelim@gmail.com}
\orcid{0000-0002-8399-1633}
\affiliation{
  \institution{Singapore University of Technology and Design}
  \streetaddress{8 Somapah Rd}
  \country{Singapore}
  \postcode{487372}
}

\author{Simon T. Perrault}
\email{perrault.simon@gmail.com}
\orcid{0000-0002-3105-9350}
\affiliation{
  \institution{Singapore University of Technology and Design}
  \streetaddress{8 Somapah Rd}
  \country{Singapore}
  \postcode{487372}
}

\renewcommand{\shortauthors}{Lim and Perrault}

\begin{abstract}
Fake news is a prevalent problem, particularly in digital media, that undermines trust and cooperation among people. As a variety of global mitigation efforts arise, the understanding of how people consider fake news becomes important, especially in local contexts. To that end, we carried out a survey with 75 participants in Singapore to understand people's perceptions of and practices with news (real and fake). Locally, fake news was found to be more pervasive in instant messaging apps than in social media, with the problem attributed more strongly to sharing than to creation. Good news sharing practices were generally observed. Highest trust was reported in government communication platforms across 11 media items. These results show that Singapore possesses a peculiar sociocultural scene, suggesting that efforts directed towards locally relevant measures may be more effective in addressing fake news in Singapore. We detail the survey results and recommended directions in this paper.
\end{abstract}




\maketitle

\section{Introduction}
While fake news is not a new phenomenon, the 2016 US presidential election brought the issue to immediate global attention with the discovery that fake news campaigns on social media had been made to influence the election~\cite{Allcott2017}. The creation and dissemination of fake news is motivated by political and financial gains, and its influence has led to increasing social costs due to the adverse effects it has on people's truth discernment and behavior~\cite{Duffy2020}. With fake news stemming mainly from digital media and causing misguided dissent that could compromise collaboration among people, we see this to be of concern to the CSCW community. As global efforts addressing fake news take off, we aim to understand what the perceptions and practices of news sharing and fake news are in a local context, with Singapore as the place of interest, to gain insights on where best to direct local mitigation efforts.

\section{Background and Related Work}

\paragraph{Fake News.} Fake news is news articles that are ``either wholly false or containing deliberately misleading elements incorporated within its content or context"~\cite{Bakir2018}. The presence of fake news has become more prolific on the Internet due to the ease of production and dissemination of information online~\cite{Shu2017}. The usage of fake news ranges from self-serving purposes like clickbait for moneymaking~\cite{Geckil2018} to agendas on a national scale like political manipulation~\cite{Allcott2017} and terrorism~\cite{Fang2021}. With the rapid and extensive adoption of social platforms, fake news has come to be more closely integrated with daily life, resulting in rising social costs due to people making poorly justified and unwarranted choices based on inaccurate knowledge~\cite{Duffy2020}. This has spurred CSCW research on areas like attitudes towards news~\cite{Wang2013}, news transmission~\cite{Liao2013}, and forms of innovative countermeasures~\cite{Bhuiyan2018, Mitra2017}, revealing the breadth of interests in this issue.

\paragraph{Fake News in Singapore.} Singapore is a city-state with an open economy and diverse population that shapes it to be an attractive and vulnerable target for fake news campaigns~\cite{Lim2019}. As a measure against fake news, the Protection from Online Falsehoods and Manipulation Act (POFMA) was passed on May 8, 2019, to empower the Singapore Government to more directly address falsehoods that hurt the public interest. The rising attention of fake news in the local scene has motivated various research including studies on the perceptions and motivations of fake news sharing~\cite{Chen2015} and responses to fake news~\cite{Tandoc2020}. Although there are parallels between these studies and ours, we want to highlight that our study explores fake news in general media instead of solely social media, examining both usage and trust. Furthermore, we investigate more broadly the attitudes and behaviors on news sharing and fake news.

\section{Method}
In this study, we seek to answer these research questions. RQ1: How much do people trust the media by which they obtain news? RQ2: Why do people share news and how do they do it? RQ3: How do people view the fake news phenomenon and what measures do they take against it? An online survey was employed for data collection in which the assessed media items include 11 forms of traditional and digital media. The survey contained 19 question items, 2 branching questions, and 3 demographic questions. Respondents were allowed to select multiple options for some question items while the branching questions served to direct them to different sections based on their answer. Although branching means that respondents may skip certain sections such that the sample for those sections become smaller, each answered question is assured to be based on the respondent’s experience rather than their estimation, which enables us to obtain more relevant and reliable self-reported data. This research was approved by the Institutional Review Board of our university.

\subsection{Procedure}
The survey was written in English and made available to anyone with the hyperlink. Participation was fully voluntary. For dissemination, various channels were employed including a mailing list of students from a local Singapore university, an informal Telegram supergroup joined by students, alumni, and faculty of the same university, and personal contacts of the researchers. Further spreading of the survey by participants was encouraged. In total, 104 responses were received.

\subsection{Participants and Data}
75 of the 104 responses fulfilled the criterion of having respondents who were currently based in Singapore. This set was extracted for further analysis and will be henceforth referred to as `SG-75'. The details on the participant demographics of SG-75 are shown in Table~\ref{tab:demo}. From SG-75, two more subsets were formed via the branching questions. The first contains 59 responses in which respondents said that they have shared news before (referred to as `SharedNews-59'), and the second contains 57 responses in which respondents said that they have come across fake news before (referred to as `SeenFake-57'). While these subsets have smaller samples, the self-reported data of the questions falling within the sections of these subsets would be more reliable since the respondents have prior experience to relate to.

\begin{table}[!htb]
  \caption{Summary of the participant demographics of the 75 Singapore-based respondents.}
  \label{tab:demo}
  \begin{tabular}{ |c|c|c|c|c|c| }
  \hline
  \multicolumn{2}{|c|}{Gender} & \multicolumn{2}{|c|}{Age} & \multicolumn{2}{|c|}{Occupation} \\
  \hline
  Male & 47 & 18-24 y/o & 48 & Students & 58\\
  \hline
  Female & 28 & 25-34 y/o & 17 & Employed & 13 \\
  \hline
  \multicolumn{2}{|c|}{} & 35-44 y/o & 7 & Others & 4 \\
  \hline
  \multicolumn{2}{|c|}{} & $>$45 y/o & 3 & \multicolumn{2}{|c|}{} \\
  \hline
  \end{tabular}
\end{table}

\section{Findings}

\begin{figure*}[!htb]
  \centering
  \includegraphics[width=.85\linewidth]{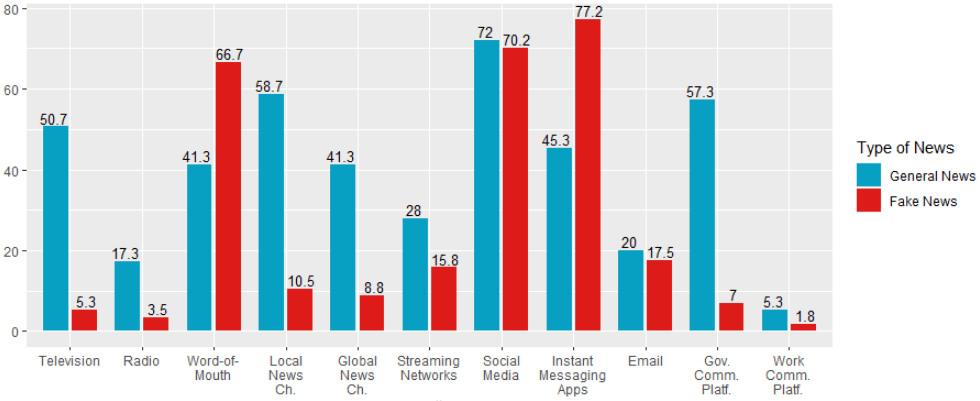}
  \caption{Reported sources of general news (out of 75 responses) and fake news (out of 57 responses) in 11 different media items. Described in percent values.}
  \label{fig:newssource}
\end{figure*}

\paragraph{News Sources (RQ1).} Respondents were asked to report the sources where they obtained news in general, and the sources where they encountered fake news based on 11 media items (see Figure~\ref{fig:newssource}). From SG-75 ($N=75$), the top three sources used to obtain news were social media ($72.0\%, n=54$), local news channels in print and digital format ($58.7\%, n=44$) and government communication platforms such as government websites ($57.3\%, n=43$). From SeenFake-57 ($N=57$), comprising respondents who have come across fake news before, the top three sources where fake news was observed were instant messaging apps ($77.2\%, n=44$), social media ($70.2\%, n=40$) and word-of-mouth ($66.7\%, n=38$).

\begin{figure*}[!htb]
  \centering
  \includegraphics[width=.85\linewidth]{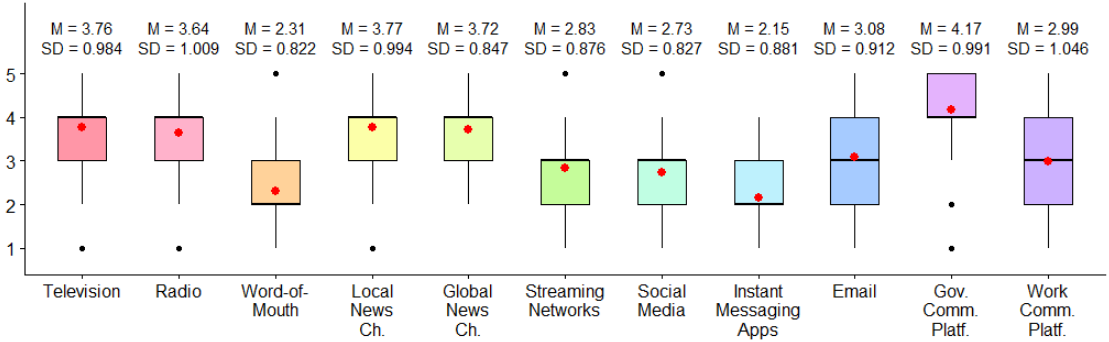}
  \caption{Reported levels of trust in 11 different media items on a 1-5 Likert scale (1: strongly distrust, 5: strongly trust). Red dots indicate the average score.}
  \label{fig:trust}
\end{figure*}

\paragraph{Trust in News Sources (RQ1).} Respondents (in SG-75) reported their level of trust in the 11 media items on a 1-5 Likert scale (see Figure~\ref{fig:trust}). A significant main effect of media on the level of trust was found using Friedman’s test (\(\chi^2(10)=338\), $p<.0001$) . Despite the popularity of social media as a general news source, respondents tended to have a low level of trust in them ($M=2.73, SD=0.827$) which could be attributed to them also being perceived as a great source of fake news (ref. Figure~\ref{fig:newssource}). Instant messaging apps were considered the least trustworthy ($M=2.15, SD=0.881$), aligning with them being viewed as the greatest source of fake news (ref. Figure~\ref{fig:newssource}). The television ($M=3.76, SD=0.984$) and local news channels ($M=3.77, SD=0.994$) held higher degrees of trust, with government communication platforms attaining the highest trust ($M=4.17, SD=0.991$) among the 11 media items.

\paragraph{On News Sharing (RQ2).} Based on SharedNews-59 ($N=59$), respondents were motivated to share news that they felt were relevant to the receiver ($89.8\%, n=53$), or were important ($83.1\%, n=49$) or interesting ($83.1\%, n=49$) to know about. News were shared largely to friends ($93.2\%, n=55$) and family ($66.1\%, n=39$), where locally popular social communication apps such as WhatsApp ($79.7\%, n=47$), Telegram ($62.7\%, n=37$), and Facebook ($33.9\%, n=20$) were the more common mode of sharing. Most respondents practiced good news sharing habits by either always ($25.4\%, n=15$) or frequently ($47.5\%, n=28$) verifying news before sharing. They do so mainly by checking and cross-referencing the news with other results from the search engine ($89.8\%, n=53$) and with information from official government channels ($88.1\%, n=52$).

\paragraph{On Fake News (RQ3).} Based on SeenFake-57 ($N=57$), we reported earlier that the top three sources of fake news were instant messaging apps, social media and word-of-mouth. Here, we detail the practices of managing fake news. In asking respondents to recall their experience when they realized that a piece of news was fake, respondents cited the search engine as the most common tool by which they ascertained the falsity of the news ($86.0\%, n=49$). Most respondents realized that the news was fake either very soon ($33.3\%, n=19$) or within the day ($42.1\%, n=24$). A mix of responses were given regarding the actions taken upon encountering fake news. They include sending a correction message to those whom they have shared the news with ($50.9\%, n=29$), sending a warning message to inform others of the presence of the fake news ($40.4\%, n=23$), or simply not doing anything ($43.9\%, n=25$).

A series of 1-5 Likert scale questions (1: strongly disagree, 5: strongly agree) were presented to the respondents (in SeenFake-57) to further gain insights into their views on fake news. Respondents feel that the issue of fake news will remain for a long time ($M=4.33, SD=0.831$), finding it unacceptable ($M=4.02, SD=1.98$) and harmful ($M=4.42, SD=0.800$). They believe it is crucial to prevent or stop fake news ($M=4.25, SD=0.931$), and that there are efforts to do so ($M=3.68, SD=0.783$), although the effectiveness of the efforts ($M=2.82, SD=0.869$) are somewhat limited. The sharing of fake news ($M=3.93, SD=1.19$) was felt to be a greater problem than its creation ($M=3.54, SD=1.20$). On personal measures, they feel that identifying fake news at a glance ($M=2.39, SD=0.861$) is less easy, while verification by checking ($M=3.44, SD=0.907$) is easier. Interestingly, the awareness of fake news among the general populace ($M=2.74, SD=1.11$) is considered to be somewhat low. Children and the elderly are also thought to be more susceptible ($M=3.72, SD=1.10$). In general, trust in news has changed since encountering fake news, with lower levels being reported (before: $M=3.56, SD=0.780$; after: $M=2.88, SD=0.709$).

\section{Discussion}
Beyond understanding the perceptions and practices of news sharing and fake news in Singapore, our results highlight opportunities in various areas that call for more locally relevant efforts in addressing fake news.

\paragraph{Improve Instant Messaging Apps.} Many studies worldwide have observed the proliferation of fake news on social media and instant messaging apps, with social media being the more commonly studied medium. In Singapore, however, mitigation efforts on fake news in instant messaging apps may be more important. Most respondents encountered fake news on instant messaging apps compared to social media, and have reported the least trust in them. They have also rated the sharing of fake news to be a greater problem than its creation. These suggest that, in Singapore, communication with personal contacts such as through the forwarding of messages, rather than with the public such as by sharing posts on social media feeds, is the larger issue. As an Asian country, Singapore tends towards a collectivist culture where emphasis is placed on establishing and maintaining relationships in one's social group. Research has shown that this is linked to lesser use of social media~\cite{Jackson2013}, and stronger preferences towards group chats in instant messaging apps~\cite{Li2011}, signaling that instant messaging apps feature more prominently in daily communication. An opportunity here is to design more effective interventions, such as warning mechanisms~\cite{Gao2018}, to preempt the private sharing of fake news.

\paragraph{Consider the Factors of Trust.} There is a very strong, negative correlation between the media sources of fake news and the level of trust in them (ref. Figures ~\ref{fig:newssource} and ~\ref{fig:trust}) which is statistically significant (\(r(9)=-0.81\), $p<.005$). Trust is built on transparency and truthfulness, and the presence of fake news, which is deceptive and usually meant to serve hidden agendas, may erode trust. It is worthwhile to consider whether the trust in media items is due to people's own encounters with fake news, or because of secondary factors. In Singapore, there have been active efforts through campaigns from various organizations (e.g., S.U.R.E.~\cite{NLB}, Better Internet~\cite{MLC}, VacciNationSG~\cite{Lai2021}) to raise awareness on misinformation, disinformation and fake news. If it is through the exposure to the messages of these campaigns that people's trust in media items have been influenced, especially those who might not have personally encountered fake news, this suggests the importance of media literacy education in addressing fake news, particularly when secondary effects such as practicing greater caution due to a lack of trust comes into play.

\paragraph{Promote Citizen Engagement by the Government.} In general, respondents possess a competent level of digital literacy skills with a majority exercising good news sharing practices. They actively verify news before sharing by checking with multiple sources found through the search engine and with authoritative information found in government communication platforms, and post corrections and warnings when they encounter fake news. That respondents show strong trust and reliance on government communication platforms, such as official websites and hotlines, signifies the relatively strong faith that Singapore residents have in the Singapore Government to provide truthful and helpful information and to debunk fake news. This may be attributed to the successful ongoing efforts in making transparent government decisions and the readiness of the government in addressing public concerns through online forums and dialogues~\cite{REACH}. There is opportunity here for the government to launch programs such as campaigns, call-to-actions and civic tech initiatives that aim to more actively involve the public in discussing the local impacts of fake news and the strategies to manage it, and to encourage them to play a part through personal and community actions.

\section{Conclusion}
In this paper, we detail the perceptions and practices of news sharing and fake news in Singapore, and discuss the opportunities for more locally relevant efforts in mitigating fake news. On the whole, trust in news has fallen. Respondents view fake news and its effects negatively and are reserved in claiming the effectiveness of current mitigation efforts, indicating that more can be done. It is important to note that the majority demographic of this study is the younger population, and that given the survey is conducted online, all participants would be somewhat digitally proficient. This thus motivates further research into the older or less digitally savvy population who may be more susceptible to fake news transmitted person-to-person such as via word-of-mouth or text messaging. Furthermore, it will be of interest to see what news people consider as `fake' in Singapore, especially given the phenomenon in the US where the term is increasingly used simply for news that one disagrees with~\cite{Nielsen2017}. Lastly, for a more robust investigation, a natural methodological progression is to incorporate interview strategies to inform the survey results in future work. With fake news posing a problem to the society at large, efforts in improving instant messaging apps, enhancing media literacy, and promoting citizen engagement to address fake news in Singapore will be worthwhile.

\bibliographystyle{ACM-Reference-Format}
\bibliography{main}

\end{document}